\documentclass[article]{aastex}

\newcommand{\myemail}{massimo.dotti@mib.infn.it}
\def\lsim{\mathrel{\rlap{\lower 3pt\hbox{$\sim$}}\raise 2.0pt\hbox{$<$}}}
\def\gsim{\mathrel{\rlap{\lower 3pt\hbox{$\sim$}}\raise 2.0pt\hbox{$>$}}}
\def\msun{\rm M_{\odot}}

\shorttitle{Massive black hole binaries}
\shortauthors{Dotti et al.}

\begin{document}

\title{Massive black hole binaries: \\ dynamical evolution and
observational signatures}

\author{M. Dotti\altaffilmark{1}}
\affil{Universit\`a di Milano Bicocca, Dipartimento di Fisica G. Occhialini, Piazza della Scienza 3, I-20126, Milano, Italy}
\email{\myemail}

\author{A. Sesana\altaffilmark{2}}
\affil{Max-Planck-Institut f\"ur Gravitationsphysik, Albert Einstein Institut, Golm, Germany}

\and

\author{R. Decarli\altaffilmark{3}}
\affil{Max-Planck-Institut f\"ur Astronomie, K\"onigstuhl 17, 69117 Heidelberg, Germany}

\begin{abstract} The study of the dynamical evolution of massive black
  hole pairs in mergers is crucial in the context of a hierarchical
  galaxy formation scenario.  The timescales for the formation and the
  coalescence of black hole binaries are still poorly constrained,
  resulting in large uncertainties in the expected rate of massive
  black hole binaries detectable in the electromagnetic and
  gravitational wave spectra. Here we review the current theoretical
  understanding of the black hole pairing in galaxy mergers, with a
  particular attention to recent developments and open issues. We
  conclude with a review of the expected observational signatures of
  massive binaries, and of the candidates discussed in literature to
  date.

\end{abstract}



\section{Introduction}

Understanding the formation and evolution of massive black holes
(MBHs) is one of the most exciting goals of contemporary astrophysics
and cosmology. It is now well established that MBHs are ubiquitous in
nearby spheroids \citep[e.g.][]{ff05}, most of them lurking in a
quiescent accretion state, while during the cosmic history
million-to-billion solar mass MBHs powered quasars. These objects have
become in the last years central building blocks for all the proposed
scenarios of galaxy formation \citep[e.g.][]{kh00}, playing a major
role in shaping galaxies through feedback processes
\citep{springel05}.  Massive black hole binaries (MBHBs), formed
during the galaxy merging process \citep{begelman80}, promise to be
among the most luminous gravitational wave (GW) sources for future
space-borne interferometers like the proposed New Gravitational wave
Observatory (NGO)
{\footnote{https://lisa-light.aei.mpg.de/bin/view/}}, and ongoing
Pulsar Timing Array (PTA) campaigns \citep{hobbs10}. Theoretical
modeling of MBHB dynamics is essential in addressing a number of
fundamental astrophysical questions (such as the merger-quasar
connection or the MBH-host relations), and in identifying putative
signatures that my serve as a guidance for present and future
observational campaigns.

Early stages of MBH pairing have been observed,
from the initial phases of galaxy mergers, where two distinct but
gravitationally bound galaxies are observable at separations of $\sim
100$ kpc \citep{hennawi,myers07,myers08,foreman,shen,farina}, down to
unbound pairs of MBHs at separations of $\lsim 1$ kpc embedded in a
single galaxy remnant \citep{komossa,colpi}.

During this initial stage the MBH pairing is driven by dynamical
friction acting on the host galaxies. The two MBHs (hereafter $M_1$
and $M_2$ for the primary and secondary MBH, respectively) bind in a
binary if they reach a relative separation
\begin{equation}\label{a_binary} a_{\rm BHB}\sim {GM_{\rm BHB} \over
2\sigma^2}\sim 0.2 \; M_{\rm
    BHB,6} \; \sigma_{100}^{-2}\;{\rm pc}, 
\end{equation} 
where $\sigma$ is the velocity dispersion of the host galaxy, $M_{\rm
BHB}=M_1+M_2$ is the total mass of the binary, and $\sigma_{100}$ and
$M_{\rm BHB,6}$ are in units of 100 km s$^{-1}$ and $10^6~ \msun$,
respectively. If, in galaxy mergers, the mass of the MBHB scales
with $\sigma$ following the MBH mass vs. $\sigma$ relation \citep[see,
e.g.][and references therein]{gultekin,graham},
equation~\ref{a_binary} implies $a_{\rm BHB}\sim 0.5 \; M_{\rm
BHB,6}^{1/2}\;{\rm pc}$.

The efficiency of the process depends on the mass ratio between the
merging galaxies. While equal mass (major) mergers result in a fast
formation of a MBHB, in very unequal mass (minor) mergers the
satellite, tidally disrupted along the course of the encounter, leaves
its {\it naked} MBH wandering in the outskirts of the most massive
galaxy \citep[see, e.g.][]{governato}. Recent numerical studies have
addressed the efficiency of the formation of a MBHB as a function of
the mass ratio between the two galaxies, the gas fraction in the
merging galaxies, and the redshift of the merger
\citep{callegari09}. In particular, the presence of a significant gas
component in the satellite helps the MBHB formation: during the first
pericenters, the interaction between the two galaxies promotes the
formation of bars, that convey a large fraction of the available gas
in the center of the merging galaxies \citep[already noticed in lower
  resolution merger simulations, see, e.g.][]{barnes96,barnes02}. This
new nuclear gas overdensity deepens the potential well of the
secondary nucleus and prevents its tidal disruption.
\citet{callegari09} found a critical galaxy mass ratio for the
formation of a MBHB of $\sim 1/10$ for gas rich galaxies and $\gsim
1/4$ for gas poor galaxies. In zero-order approximation, assuming the
MBH mass vs. bulge mass relation \citep{haring,marconi} we expect
similar mass ratios $q=M_2/M_1\; (\le 1)$ in MBHBs. However, $q$ can
be increased for gas rich galaxies. The strong gas inflows in the
satellite (more perturbed) galaxy can result in a faster growth of the
smaller MBH, and in higher values of the expected MBH mass ratio
\citep[$q\gsim1/3$][]{callegari11}\footnote{We caution that these
  ranges in $q$ implicitly assume similar morphology in the two
  merging galaxies, although minor mergers can involve very different
  morphological types. A study of MBH pairing and the formation of
  MBHBs in mixed mergers (e.g. disks merging with ellipticals) is not
  available to date.}.

As a consequence, efficient dynamical friction promotes the formation
of MBH binaries with similar mass ratios. The expected number of
observable binaries, and the rate of MBH coalescences, however,
depends on the dynamical evolution after the binary formation. In
order to coalesce through GW emission in less
than an Hubble time, the two MBHs have to reach a separation
\begin{equation}\label{eq:agw} a_{\rm GW} \approx 2\times 10^{-3}
f(e)^{1/4}{q^{1/4}\over(1+q)^{1/2}}\left ({M_{\rm BHB} \over
  10^6\,\msun}\right )^{3/4}\,{\rm pc},
\end{equation}
where $f(e)=[1+(73/24)e^2+(37/96)e^4](1-e^2)^{-7/2}$ is a function of
the binary eccentricity $e$ \citep[][]{peters64}.
Circular equal mass binaries can coalesce  shrinking by a factor
\begin{equation}
\frac{a_{\rm BHB}}{a_{\rm GW}}\sim 350 \left ({M_{\rm BHB} \over
  10^6\,\msun}\right )^{-1/4},
\end{equation}
while this factor decreases to 1 for $10^6\,\msun$ binary with $e
\approx 0.999$. In order to understand the final fate of a MBHB and to
constrain theoretically its observability, it is fundamental to study
at the same time the orbital decay and the eccentricity evolution of
the binary. The final fate of the MBHs, i.e. if they will coalesce in
a single object or not, strongly depends on the amount of matter
(stars and gas) they can interact with after the binary formation. A
definite answer is not present to date. In this paper different
scenarios will be discussed, depending on the nature of the galaxy
mergers and on the properties of the galaxy nuclei.

This paper is organized as follow: In Section~\ref{gaspoor} we review
the dynamical evolution of MBHB in gas poor environments, while the
effect of gas is discussed in Section~\ref{gasrich}. In
Section~\ref{roris} we describe the MBHB candidates observed to
date. Finally, our conclusions are drawn in Section~\ref{conclusions}.

\section{Dynamical evolution in gas-poor environment}\label{gaspoor}
In systems where dynamical friction is efficient in dragging the
two MBHs to the center of the merger remnant, the now bound MBHB 
is inevitably embedded in a gas and star rich environment. Such
rich ambient provides a variety of physical mechanisms to efficiently
extract the energy and angular momentum of the MBHB, promoting its
final coalescence. In this section, we focus on 
dynamical processes involving interactions with stars. MBHBs in
pure stellar environments were the first to be examined \citep[the 
basics going back to][]{begelman80}, for the obvious reason that stars 
can be considered as point particles, affected by gravitational 
forces only. The MBHB-star interactions are therefore adequately 
described by Newton's laws only, without all the complications 
involved in gas dynamics. Nonetheless, a single star-binary interaction
is, by definition, a three body problem, and the dynamics of the
system is inevitably chaotic. Therefore, no simple analytical 
solutions are viable, and numerical studies (both involving three 
body scatterings and full N-body simulations) have been massively 
exploited to tackle the problem. The fate of the MBHB is 
determined by its semimajor axis and eccentricity evolution
(see the introduction); in the following we discuss them separately. 

\subsection{Shrinking of the binary semimajor axis}
The basic physical process driving the MBHB evolution in presence of 
stars is the slingshot mechanism. A star intersecting the MBHB orbit
undergoes a complex three-body interaction being eventually 
ejected at infinity, carrying away energy and angular momentum 
from the binary. Extensive three body scattering experiments
\citep{mv92,qui96,shm06} have shown that ejected stars carry away
an energy per unit mass of the order of 
\begin{equation}
\Delta{E_{\rm BHB}}\approx \frac{3}{2}\frac{G\mu}{a_{\rm BHB}},
\end{equation} 
where $\mu=M_1M_2/M_{\rm BHB}$ is the reduced mass of the binary.
Assuming a classical interaction rate given by 
$\Gamma=\rho/m\Sigma v$, where $\Sigma$ is the binary cross section, 
$\rho/m=n$ is the number density 
of the ambient stars, and $v$ is their typical velocity 'at infinity' 
with respect to the binary (i.e. the velocity dispersion $\sigma$ of 
the stellar system), \cite{qui96} showed that the evolution of the 
binary semimajor axis is simply given as 
\begin{equation}
\frac{da_{\rm BHB}}{dt}= -\frac{a_{\rm BHB}^2G\rho}{\sigma}H,
\label{hrate}
\end{equation} 
where $H$ is a dimensionless {\it hardening rate}. If 
$a_{\rm BHB}<GM_2/(4\sigma^2)$, $H\approx 15$ independently on the mass, 
mass ratio and eccentricity of the system. In principle, given a
stellar system with density $\rho$ and velocity dispersion $\sigma$, equation 
(\ref{hrate}) predicts efficient coalescence of the MBHB.

However, the above simple treatment ignores the concept of loss
cone depletion. In an extended stellar system, only a tiny fraction
of the stellar phase space allows orbits intersecting the MBHB, 
commonly referred as 'binary loss cone'. As stars are ejected, the
loss cone is depleted, and the binary evolution is governed by
the rate at which new stars are fed into the loss cone \citep{mf04}. 
In typical stellar systems, the mass in stars in the binary loss cone is 
of the order of few times $\mu$ \citep{mm05}, insufficient to reach $a_{gw}$
in most of the cases \citep{shm07}. This is the origin of the 
'last parsec problem' \citep{mm01}. In a spherical stellar
system, the loss cone refilling proceeds on a two body relaxation
timescale \citep{bt87}, which is usually much longer than the 
Hubble time. In the last decade, this fact has been confirmed in N-body 
simulations \citep{mm01,mf04,mms07}. In such simulations, after loss cone
depletion, further hardening was provided by two body relaxation.
This is a process that depends on the 'granularity' of the systems,
and the result is an N-dependent hardening rate, with the binary 
evolution slowing down as the number of particle in the simulation 
increases. Extrapolating these results to a realistic N representative 
of a galactic bulge, the binary evolution would have stalled. 

In recent years, evidence has emerged that the 'last parsec
problem' might be an artificial product of the 'spherical cow'
approximation which is often exploited in astronomy. Basically,
the spherical systems studied in the simulations represent a 
worse case (and unrealistic) scenario. MBHBs are infact produced
in galaxy mergers, in which the resulting stellar bulge is 
rotating, triaxial and likely to undergo bar-like 
instabilities. In a triaxial potential, an orbiting star
does not conserve any of its angular momentum components 
\citep{bt87}. As a result, there is a vast family of orbits 
(called centrophilic) that are allowed to get arbitrarily close 
to the binary \citep{pm04,mp04,mv11}, keeping the loss cone full during
the MBHB hardening process. Recent N-body simulations 
have confirmed this scenario. \cite{ber06} studied 
the evolution of a MBHB in a rotating bulge. In this case,
the stellar system experiences a bar instability resulting 
in a triaxial potential.
The binary hardening rate was found to be N-independent; a 
proof that the hardening was not proceeding because of spurious
two body relaxation. More recently, the advent of GPU computing
made possible to simulate 'ab initio' the evolution of two 
interacting stellar bulges hosting MBHs; a first step toward
a realistic galactic merger scenario. 
Several simulations were performed by \cite{khan11}, \cite{preto11}
and \cite{mg11}.
In all cases, the stellar remnant was triaxial and rotating,
and the hardening rate, given by triaxiality driven loss cone 
replenishment was found to be independent on N, implying coalescence 
timescales of $\approx10^8$yr. Remarkably, when normalized to the 
merging galaxy properties, the binary hardening rates found 
in these simulations follow equation (\ref{hrate}) where
$H\approx20$ \citep{mg11}. This is a consequence of the fact that,
whatever is the geometry of the system, the average 'quantum' 
of energy taken away from an interacting star is always the same,
and the evolution of the system is determined by the star-binary
interaction rate only. 

\subsection{Eccentricity evolution}\label{sec:eccstars}
As pointed out in the introduction, eccentricity plays an important role
in driving the binary coalescence. However, addressing the 
eccentricity evolution of the system is more complicated because
the $\Delta e_{\rm BHB}$ caused by each individual interaction depends
on a combination of energy and angular momentum exchanges. The
angular momentum distribution of the interacting stars is 
therefore crucial. The eccentricity evolution can be described as 
\begin{equation}
\frac{de_{\rm BHB}}{d{\rm ln}(1/a_{\rm BHB})}=K
\end{equation}
Here $K$ is a dimensionless parameter that, differently than
$H$, depends on the binary mass ratio, 
and eccentricity itself \citep{qui96,shm06}. 
In general $K$ is a positive number in the range $0-0.3$
(the peak value occours at $e_{\rm BHB}\approx 0.6$), 
meaning that the binary eccentricity grows during the shrinking process.
\cite{sesana10} constructed complete binary evolutionary tracks
by coupling three body scattering experiments of bound and 
unbound stars to an analytical description of the stellar 
distribution and on the loss cone refilling. As a general
trend, quasi circular-equal mass MBHBs experience just a mild
eccentricity growth, while systems which are already eccentric
at the moment of pairing, or with $q$ significantly lower than
1, can evolve up to $e_{\rm BHB}>0.9$.

The eccentricity evolution in stellar environments has been tackled by
several authors by means of full N-body simulations. However, the
limited number of particles ($N<10^6$) in such simulations results in
very noisy behavior for the binary eccentricity, and it is difficult
to draw conclusions about the general trends behind the numerical
noise. \cite{mm01} carried out numerical integration of equal MBHBs
embedded in two merging isothermal cusps ($\rho\propto r^{-\gamma}$, 
with $\gamma=2$). Starting with
circular orbits they find a mild eccentricity increase to a value of
$\lsim0.2$ during the stellar driven hardening phase. \cite{mms07}
considered equal MBHBs embedded in Dehnen density profiles
\citep{de93} with $\gamma=1.2$ with different initial eccentricities.
Again, they find that circular binaries tend to stay circular, while
eccentric binaries tend to increase their eccentricities in reasonable
agreement with the prediction of scattering experiments.  Simulations
starting before the formation of MBHBs carried by \cite{hm02} and
\cite{ar03} produce binaries with $e_0\approx 0.8$ at the moment of
pairing, with $e$ subsequently increasing up to $\gsim0.95$.
\cite{as09} focused on binaries of intermediate MBHs
($M\sim10^3\msun$) in massive star clusters. Coupling full N-body
simulations to three body scattering experiments, they find binaries
with significant eccentricity ($\sim0.5-0.6$) at the moment of
pairing, growing up to $>0.9$ during the hardening phase; similar
conclusions are reported by \cite{as10}. On the small $q$ side,
simulations were performed by \cite{bau06} and \cite{mat07}, assuming
a stellar density profile $\gamma=1.75$, motivated by the analytical 
equilibrium solution for a dense relaxed stallar cusp around a massive 
object \citep{bw76}. When properly rescaled, the
eccentricity increase found in both papers agrees remarkably well with
predictions based on the hybrid model by \cite{shm08}. \cite{iwa11}
investigated in detail the angular momentum exchanges between the
binary and the stars responsible for the eccentricity growth, in bound
stellar cusps. In particular they showed that stars counterrotating
with the binary tend to extract a lot of angular momentum from the
MBHB, causing the eccentricity growth, whereas corotating stars do
not. This is a simple consequence of angular momentum conservation
during the ejection process, as shown by \cite{sgd11}.

The evolution of the binary eccentricity can be extremely different
for non isotropic systems. For example, \cite{dotti07} showed that at
large scales before the formation of a binary dynamical friction
exerted by rotationally supported stellar disks tend to circularize
the orbit of a MBH pair. At smaller separations, \cite{sgd11}
demonstrated that in a rotating stellar system, the eccentricity
evolution of unequal MBHB is dramatically affected by the level of
co/counter rotation of the stellar distribution with respect to the
binary, with corotating distributions promoting circularization rather
than eccentricity growth. Nonetheless, most of the simulations
involving rotating bulges \citep{ber06,ber09}, or merging systems
\citep{khan11,preto11,mg11} find quite eccentric binaries at the
moment of pairing (ranging from $0.4$ to $0.8$), and the subsequent
evolution leads to a general eccentricity growth, in good agreement of
what predicted for an isotropic stellar distribution. This may be
because the binary evolution is mostly driven by loss cone refilling
of unbound stars on almost radial orbits, with negligible initial
angular momentum.

Overall, the emerging general picture favors efficient coalescence
of MBHBs in dense stellar merger remnants. The triaxial and 
rotating nature of the stellar distribution promotes efficient
loss cone refilling, while large eccentricities (especially in unequal
mass systems) shorten the gap between the binary pairing 
and the efficient GW emission stage. In the near future, 
massive N-body simulations with several million particles 
will offer a unique opportunity to confirm this scenario. 

\section{Dynamical evolution in gas-rich environment}\label{gasrich}

\subsection{Formation of a MBH binary in a circumnuclear gas disks}\label{largescale}

\begin{figure} \begin{center}
\includegraphics[width=\textwidth]{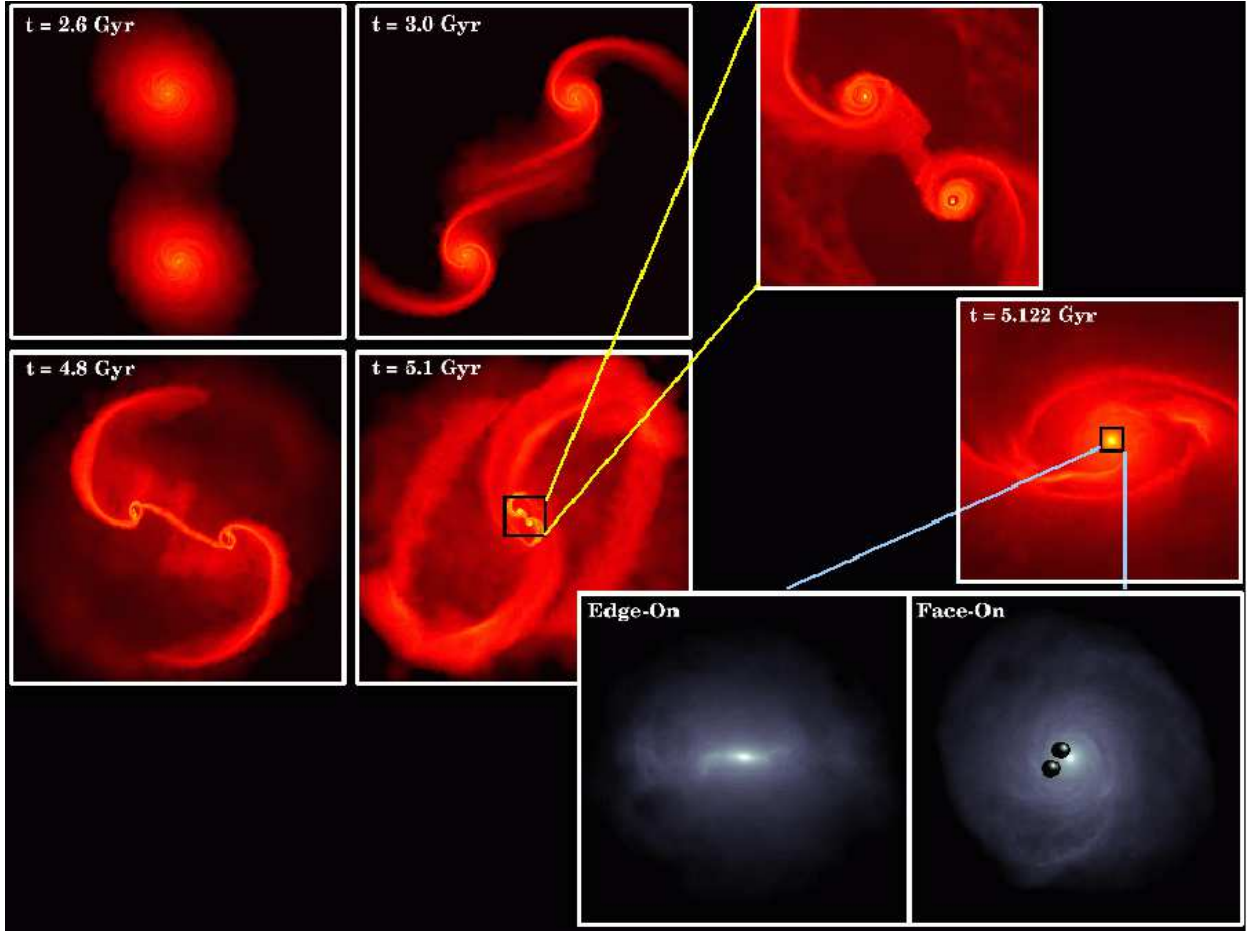}\\ \caption{Different
stages of an equal mass gas rich merger \citep[from][]{mayer07}. The
color code refers to gas density. The four panels to the left show the
large-scale evolution at different times. The strong inflow of gas
onto the two galactic centers is observable after the first
pericenter. The boxes are 120 kps on a side (top) and 60 kpc on a side
(bottom). The panels on the upper right are zoom-ins (8 kpc on a side)
of the central region during the merger of the two galactic cores. The
two bottom panels show the circumnuclear disk forming in the center of 
the galaxy remnant. The MBHs are shown as black spheres. 
}\label{lucio}
\end{center}
\end{figure}

As discussed in the Introduction, in comparable mass, gas rich galaxy
mergers the gravitational interaction drives strong inflows of gas
toward the galactic centers. The numerical multi-scale investigation
of an equal mass galaxy merger discussed in \citet{mayer07} revealed
that, in advanced stages of the galaxy merger the two MBHs, orbiting
in the central ~100 pc of the merger remnant, are embedded in a dense,
rotationally supported, gas disk (see figure~\ref{lucio}). This
{\it circumnuclear} disk is self-gravitating, and can be up to $\sim$
500 times more massive than the MBH pair \citep{mayer07}. The
dynamical evolution of the two MBHs is driven by dynamical friction,
and, since the circumnuclear disk is the densest structure in the
remnant nucleus, it is the main cause of their orbital
decay. \citet{mayer07} followed the evolution of the two MBHs from the
initial stages of the galaxy merger down to to $\lsim$ 5 pc where they
form a binary, as the mass of gas enclosed within their separation is
less than the mass of the binary.

The evolution of MBHs in circumnuclear disks has been studied in
details in dedicated simulations, in which the former evolution of the
MBHs at distances $\gsim 100$ pc is not explored. This allows to achieve
a better resolution in the central region of the remnant and to study the
latter MBH pairing. Similar indipendent investigations discussed in
\citet{escala05} and \citet{dotti07,dotti09} agree with
\citet{mayer07} on a rapid (on a timescale of $\lsim 10^6$ yr)
formation of a MBHB at parsec separations. The higher resolution in
these studies allow for a further decay of the binary, that
reaches sub-pc separations comparable to the spatial resolution in
these numerical studies.

During this intermediate stage (100 pc $\gsim a \gsim$ 0.1 pc) the
interaction with the circumnuclear disk strongly affects the
eccentricity of the MBH pair. \citet{dotti06} first noted that, for
MBHs with an initial significant eccentricity the decay phase is
preceded by a circularization of the orbit. This is due to the
dynamical friction exerted by a rotating background onto the BHs at
their apocenters. Since the circumnuclear disk is rotationally
supported, the BHs at their apocenter move slowlier than the local
gas, and are dragged in the direction of their motion. This positive
torque is exerted only in proximity of the apocenter, where the
angular momentum of the MBHs can maximally increased. This effect is a
general feature of dynamical friction exerted by a rotating
background. \citet{dotti07} demonstrated that the same effect is
present for MBHBs orbiting in a stellar circumnuclear disk (as discussed in Section~\ref{sec:eccstars}), and
\citet{callegari09} found the same effect acting in unequal mass
galaxy mergers at larger scales, where the disk of the primary galaxy
circularizes the orbit of the satellite.

\subsection{Evolution of close MBH binaries in circumbinary gas disks}

After the formation of a close binary, the MBHs, acting as a source of
angular momentum, exerts a tidal torque that inhibits the gas from
drifting inside its orbit. \begin{figure} \begin{center}
\includegraphics[width=8cm]{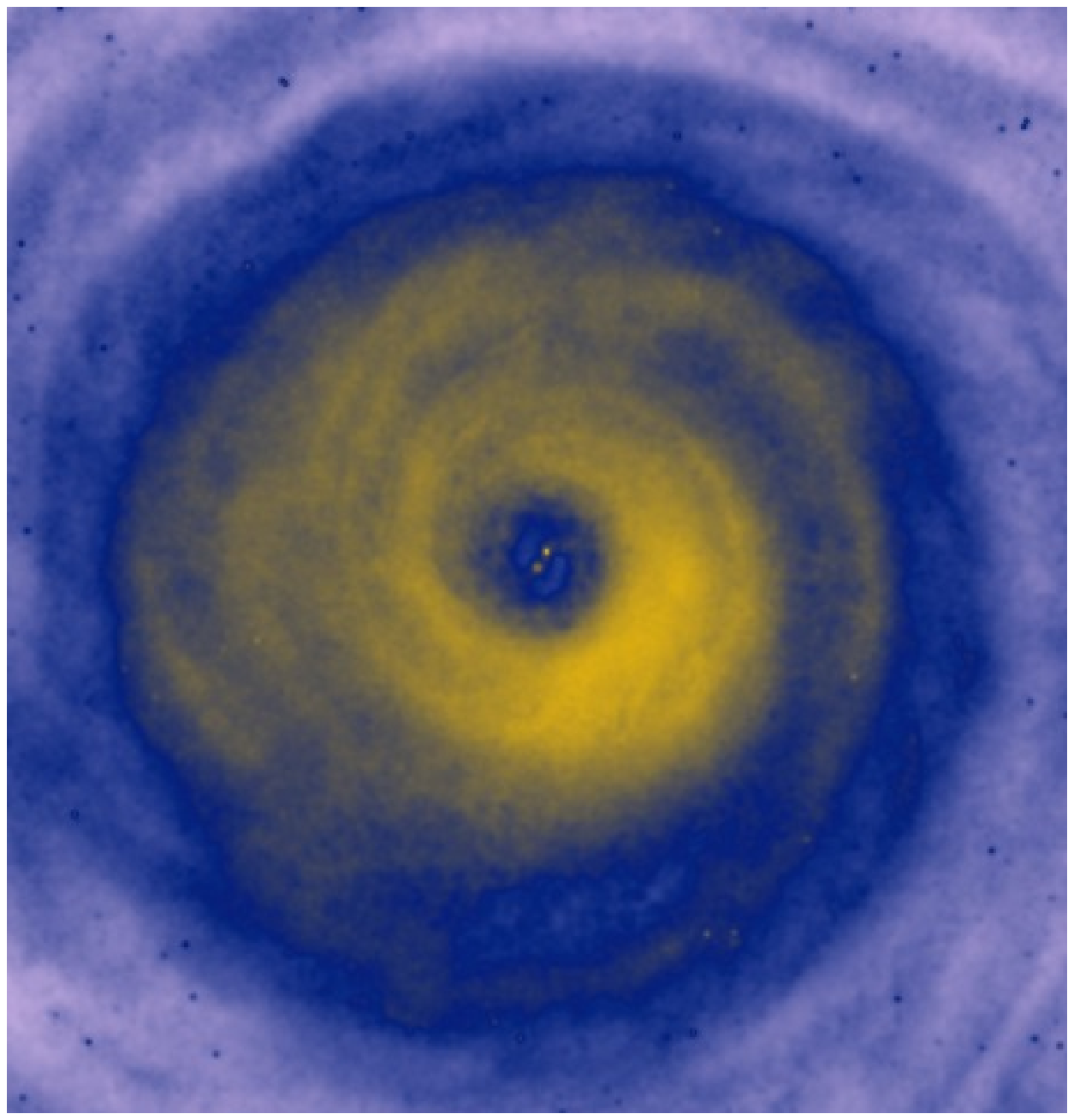}\includegraphics[width=8.2cm]{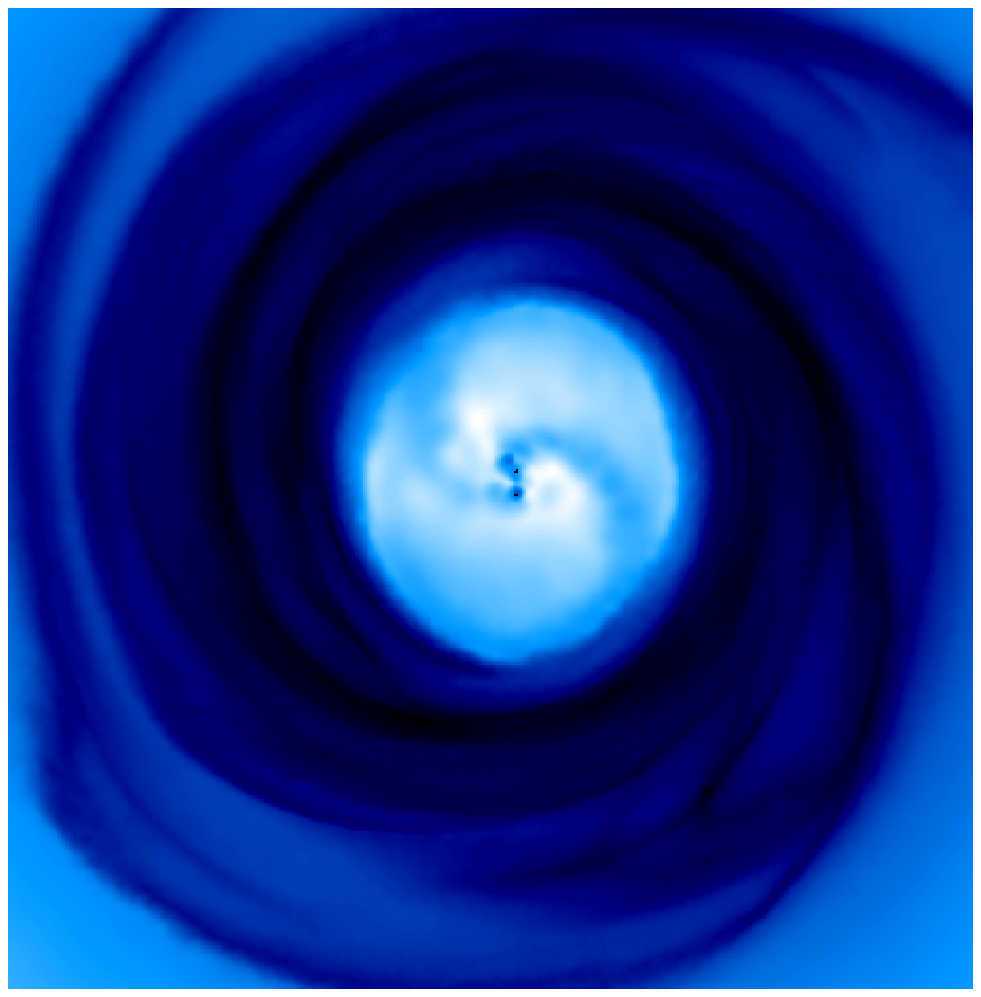}\\
\caption{Two consecutive stages of the MBHB decay in the nucleus of a
gas rich galaxy merger remnant.  Left panel: Central region ($\sim$ 5 pc) of
a large-scale (100 pc) massive ($10^8 \msun$) circumnuclear disk,
embedding a sub-pc MBHB \citep[from one of the simulations in][]{dotti09}. 
The color code refers to the gas density (high
densities in yellow). Right panel: an eccentric MBHB embedded in a
sub-pc circumbinary disk \citep[from][]{roedig11}. Here darker blue
refers to higher densities.}\label{gap}
\end{center}
\end{figure}
This creates a hollow density region,
called gap, that surrounds the binary \citep[e.g.][]{linpapaloizou79a,
  linpapaloizou79b, artymowicz94, syer95, gould00}.

As a consequence of disk clearance, corotation and inner Lindblad
resonances are reduced in power, drastically changing the dynamical
evolution of the binary. The same transfer of angular momentum that
keeps the disk from infalling onto the MBHs drives the shrinking of
the binary. Analytical and numerical studies agree in predicting that
the interaction between the binary and the surrounding material
reduces the semimajor axis of the binary, and increases the
eccentricity of an initially low eccentricity binary
\citep[e.g.][]{papaloizou77, goldreich80, ivanov99, gould00,
  armitage02, goldreich03, armitage05, haiman09, cuadra09, 
  roedig11}.

The MBHs/circumbinary disk interaction can be studied in two limit
cases: $i)$ assuming that the MBHB is embedded in a virtually infinite
disk, as in the case in which it is continuously refilled by a
long-lived larger-scale structure, or $ii)$ assuming that the disk has
a finite mass (and, as a consequence, a finite reservoire of energy
and angular momentum).

In the first limit of a constantly fuelled disk, analytical models of
the evolution of the binary are available. 
The orbital
decay timescale is:
\begin{equation} 
t_{\rm decay}=\frac{M_{\rm disk}(a)+M_2}{M_{disk}(a)} t_{\rm visc},
\end{equation} 
Where $M_{\rm disk}(a)$ is an estimate of the disk mass inside the
orbit of the secondary MBH.  As long as $M_{\rm disk}(a)$ is greater
than $M_2$, the MBH behaves as a fluid element, decaying on the
viscous timescale $t_{\rm visc}$ onto the primary. As the binary orbit
shrinks, $M_{\rm disk}(a)$ decreases. When the enclosed mass becomes
comparable with the secondary its decay timescale increases to
resulting in very large migration timescales at small separations.
However, as first noticed in \cite{ivanov99}, in a continuously
refilled disk the stalling of the binary would cause a staedy increase
of the density of the inner edge of the circumbinary disk, until the
mass close to the binary becomes again comparable to the secondary. At
this stage, fast migration starts again. Following \cite{ivanov99},
\cite{cuadra09} estimated that, for a disk on the verge to undergo
fragmentation (i.e. as dense as possible), the coalescence timescale
of a $3\times 10^6 \msun$ binary with $q=1/3$, starting from an
initial separation of $< 0.05$ pc, is $<10^9$ yr. This timescale
decreases with the decrease of the binary mass.  Promisingly, the
initial separation assumed here is close to the limit achieved in the
larger scale simulations discussed in Section~\ref{largescale}.  Note,
however, that such a timescale is comparable with the age of the
Universe at $z \gsim 7$. If the migration in a dense circumbinary disk
is the fastest process driving the MBHs coalescence, no coalescences
of binary with $M_{\rm BHB}\gsim 10^6 \msun$ are expected at $z \gsim
7$.

In a similar way, the study of the evolution of $e$ is possible.  For
simple alpha disks \citep{ss73}, \cite{artymowicz94} found that the
outer edge of the gap (i.e. the inner edge of the circumbinary disk)
depends on the binary eccentricity, with more eccentric binaries
opening larger gaps.  However, the eccentricity itself increases as a
consequence of the interaction with the expanding disk
\citep{goldreich80, goldreich03, armitage05, cuadra09, roedig11},
resulting in a steady expansion of the gap. These two effect together
would result in a limiting eccentricity of $\sim 1$, leading to a fast
coalescence due to efficient gravitational wave emissions (see
eq~\ref{eq:agw}).  A smaller limiting eccentricity may result from a
less efficient coupling between the gas and the binary, as expected if
the disk moves farther out. Since the evolution of $e$ in an expanding
gap has been performed with finite mass disks we postpone its
discussion to the following.

If the circumbinary disk is limited in mass, the evolution of the
orbital separation and eccentricity can be quite different. Since in
this scenario the disk is not continuously refilled from the outside,
the gas mass initially available in the disk is the key parameter:
\begin{itemize} 

\item If the disk is $ \gg M_2$, the evolution is similar to the
infinite-mass disk, with the binary coalescing on a short
time-scale. 

\item If the mass in the circumbinary disk is less or of the order of
$M_2$, the interaction with the binary forces the whole circumbinary
disk to move outward in few orbital periods. This expansion of the
circumnuclear disk has been observed in simulations in which the
components of the binary have similar masses
\citep[e.g.][]{cuadra09,roedig11}. In this case, the gas reaches
distances $\gsim 4 a$, at which the interaction with the binary is not
efficient anymore. A small amount of gas can then fall again closer to
the binary because of orbital angular momentum exchange with the bulk
of the disk, but most of the gas would never get close enough to
significantly alter the evolution of the binary. Note that the
expansion of the circumbinary disk is most effective for eccentric
binaries. Binaries do not coalesce because of the interaction with too
small ($\lsim M_2$), non-refuelled disks.
\end{itemize}

In this simple description, the mass in gas within the disk is either
accreted onto the MBHs or conserved. However, the gas in such a dense
environment could be consumpted by star formation, decreasing the
effective mass of the circumbinary disk. As a consequence, the disk
could be initially $\gg M_2$, but decreasing in mass with time, and
could possibly fail in bringing the binary to the final
coalescence. This scenario has been recently discussed in
\cite{lodato09}. In this investigation stars are allowed to form in
the disk whenever it becomes gravitationally unstable. The rate of new
star formation is obtained requiring that they would inject enough
energy to keep the disk on the verge of fragmentation (i.e. providing
an heating term exactly equal to the cooling losses in the disk). Even
considering such a simple "thermal" feedback from the newly formed
stars, the disk loses so much mass that the binary cannot reach the
final coalescence, unless its initial separation is $\lsim 0.01$ pc
(for the MBH masses considered in the paper, $M_1 = 10^8$ M$_\odot$
and $M_2=10^7$ M$_{\odot}$).

Other possible feedback terms, not included in this model, that can
help in preventing such a strong gas consumption has been suggested by
Lodato and collaborators, such as momentum feedback from stellar winds
and supernovae explosions. Furthermore, the interaction between the
MBHs and the forming gas clumps and stars, not considered in the
investigation, could help the binary decay. The consumption of gas
may not be a problem if large inflows of gas are present, as in the
case of a continuously refuelled disk discussed above.

In the finite mass disk scenario, the existence of a limiting
eccentricity has been studied in \cite{roedig11} through a suite of
high resolution SPH simulations. They find a critical value $e_{\rm
crit} \approx 0.6 - 0.8$. In these simulations, the initial
ratio $\delta$ between the gap size and the semimajor axis of the
binary is 2, and can increase during the runs up to more than 4, when
the interaction efficiency drops. The analytical model presented in
the paper agrees with the simulations, predicting the limiting
eccentricity to be: \begin{equation} e_{\rm crit}=0.66 \sqrt{{\rm
ln}(\delta - 0.65)}+0.19.  \end{equation}

The initial choice of $\delta = 2$ is somewhat arbitrary. In
reality, the feeding of a MBH binary forming in a gas-rich galaxy
merger can be a very dynamic process, and the interaction with a
single circumbinary disk could be too idealized a picture. Larger
scale simulations show episodic gas inflows due to the dynamical
evolution of the nucleus of the remnant \citep[see e.g.][]{escala06,
  hopkins10}. In this scenario the binary can still interact with a
disk and excavate a gap, but the size of it would be time-dependent
(as in the simulations presented here), and would depend on the
angular momentum distribution of the inflowing streams, resulting in a
range of $e_{\rm crit}$.

Note that the discussion above implicitly assumes that the MBHB and
the circumnuclear disk corotate with each other. This is the natural
outcome of a evolutionary sequence in a gas rich galaxy mergers, in
which the two MBHs orbit in a large scale circumnuclear disk, are
forced to corotate with it \citep{dotti09}, and
open a gap in the very central region of the gas distribution. This
picture, however, could not apply in gas-poorer mergers, or even in a
gas rich scenario, if the circumnuclear disk formed during the merger
fails in bringing the two MBHs to the final coalescence before
it is consumed by star formation and/or MBH feedback. In one of
these cases, an occasional small inflow of gas could happen with a
random angular momentum, and could form a retrograde circumbinary
disk, counterrotating with respect to the binary. 

The evolution of a MBHB in a retrograde disk has been discussed in
\cite{nixon11}. In this case, the gravitational interaction between
the binary and the gas brakes both the components, so that, unlike in
the prograde scenario, here the torques responsable for the binary
shrinking and its eccentricity evolution causes the edge of the disk to
move inwards. The binary, and the secondary MBH in particular,
experiences the presence of a closer distribution of gas, which would
imply a faster evolution, moving at a higher relative velocity, that
results in a less effective interaction. Nixon and collaborators show
that $i$) if the binary is initially not exactly circular
($e<e_{\rm crit}\sim H/R$, where $H/R$ is the aspect ratio of the
disk), and $ii)$ if it interacts with the disk mainly at the
apocenter, then the secondary evolves onto an almost radial orbit after
interacting with a gas mass comparable to its own \footnote{Note that
  assuming the interaction to take place mainly at the apocenter is in
  agreement with the circumbinary disk to form after the MBHs bind in
  a close binary, as a consequence of a randomly oriented accretion
  event.}. 

The increase of the eccentricity to $e \sim 1$ in the retrograde case is
due to direct accretion of linear momentum from counterrotating
material. Since the secondary has null radial velocity at the apocenter,
before and after the interaction, and far
from the apocenter the secondary is assumed to move on an unperturbed
Keplerian orbit, the apocenter is constant. Interacting with
counterrotating gas the secondary decreases its angular momentum,
reducing its semimajor axis (of up to a factor of 2) and, most
importantly, increasing its eccentricity (up to 1).
At very high eccentricities the emission of gravitational
waves can bring the binary to the final coalescence in less than an
Hubble time (see eq.~\ref{eq:agw}).

Note that this scenario suffers of the same disk consumption problem
as the prograde one. If the disk is consumed by star formation or
evacuated by MBH or supernovae feedbacks, the process stops. This
makes this process particularly interesting for very unequal mass
binaries ($M_2/M_1\lsim H/R$, less likely to form from galaxy mergers,
as discussed in the Introduction). Fast inflows of gas, on timescales
shorter than the consuption time, would help the coalescence in both
the prograde and the retrograde scenario.

\section{Binary candidates in the realm of observations}\label{roris}

Despite being a natural outcome of galaxy mergers, MBH pairs are still
elusive. Less than 20 systems with separations of $\sim10$ pc to
$\sim10$ kpc pairs are of this kind are known to date. MBHs orbit in
the common post-merger stellar environment, in-spiralling because of
dynamical friction. They appear as a single galaxy (eventually, with
disturbed morphology) with two active nuclei. Examples are the
prototypical case of NGC 6240 \citep{komossa}; the radio galaxy 3C75
\citep{hudson06}; the spiral galaxy NGC 3341 \citep{barth08}; the
ULIRG Mrk463 \citep{bianchi08}; the interacting galaxy COSMOS
J100043.15+020637.2 \citep{comerford09}; and the quasar pair
J1254+0846 \citep{green10}.  All these objects have been discovered
because of the presence of two resolved X-ray sources wandering in the
merged galaxy. In order to look for these systems, an alternative
approach is to search for objects with two systems of narrow lines at
slightly different redshifts \citep{wang09,liu10}. Large spectroscopic
surveys, like the Sloan Digital Sky Survey \citep[SDSS][]{york00},
have been used to search for these systems. Follow-up observations
were then used to discriminate between dual AGN and single AGN with
complex gas dynamics in the narrow line region \citep{wang09,fu11}.

At separations of $\lesssim$ 10 pc the two MBHs start experiencing
their own gravitational interaction, binding in a binary. These
systems cannot be spatially resolved in optical and X-ray
observations, and radio interferometry is required. This has been
successfully done only in the case of 0402+379
\citep{maness04,rodriguez06}. The two flat-spectrum radio sources,
corresponding to the two components of the candidate MBHB, have a
projected separation of $\sim$ 7 pc (few milliarcsec at $z=0.055$).
We note however that radio interferometry at very high spatial
resolution is not an efficient technique to search for rare objects
as MBHBs, because of the limited field of views and the requirement
that both the MBHs are radio-luminous \citep[see, f.i.,][]{burke11}.

Another approach to look for MBHBs is to study periodic variations
in the luminosity of some AGN. The only MBHB candidate selected on
these bases up to now is the BL Lac object OJ287 \citep[see][and
references therein]{valtonen08}. It shows a 11 yr periodic flaring.
In the MBHB scenario, the secondary MBH orbits on a tilted plane
with respect to the accrtion disk of the primary. The flares are
associated to the passages of the secondary MBH through the nodes.
However, this is not the only available explanation of the peculiar
behavior of this source \citep{villforth10}.

More promisingly, signatures of MBHBs have been searched for in
optical and NIR spectroscopic databases. According to the MBHB
hypothesis, if at least one of the MBHs is active, the broad lines
(BLs) emitted by gas bound to it may be red- or blue-shifted with
respect to their host galaxy systemic recessional velocity, as a
consequence of the Keplerian motion of the binary
\citep{begelman80}. Therefore, looking for quasars with
significant velocity shifts ($>$few hundreds km s$^{-1}$) can be a
valid approach to systematically search for MBHB candidates over
large fields. This technique does not suffer of any angular
resolution limitations: Actually, the closer (and more massive) the
binary is, the more shifted/deformed the BLs are. Five MBHB
candidates have been individually found in this way: SDSS J0927+2943
\citep{komossa08,bogdanovic09,dotti09b}, J1536+0441
\citep{boroson09}, J1050+3456 \citep{shields09}, J1000+2233
\citep{decarli10}, and J0932+0318 \citep{barrows11}. \citet{vivi11}
recently applied this technique in a systematic way over the whole
spectroscopic sample of SDSS, using a method developed for searching
unresolved gravitational lenses \citep{tsalmantza11}. This analysis
resulted in 4 new MBHB candidates (J1012+2613, J1154+0134,
J1539+3333 and J1714+3327) and the confirmation of all the
previously known objects. In a similar study, \citet{eracleous11} searched
for objects with anomalous H$\beta$ profiles in the SDSS quasar catalogue
\citep{schneider10}. Among them, they identified 88 sources showing
velocity shifts between the broad H$\beta$ line peak and the rest-frame
of the narrow emission lines. 
\begin{figure}
\begin{center}
\includegraphics[width=\textwidth]{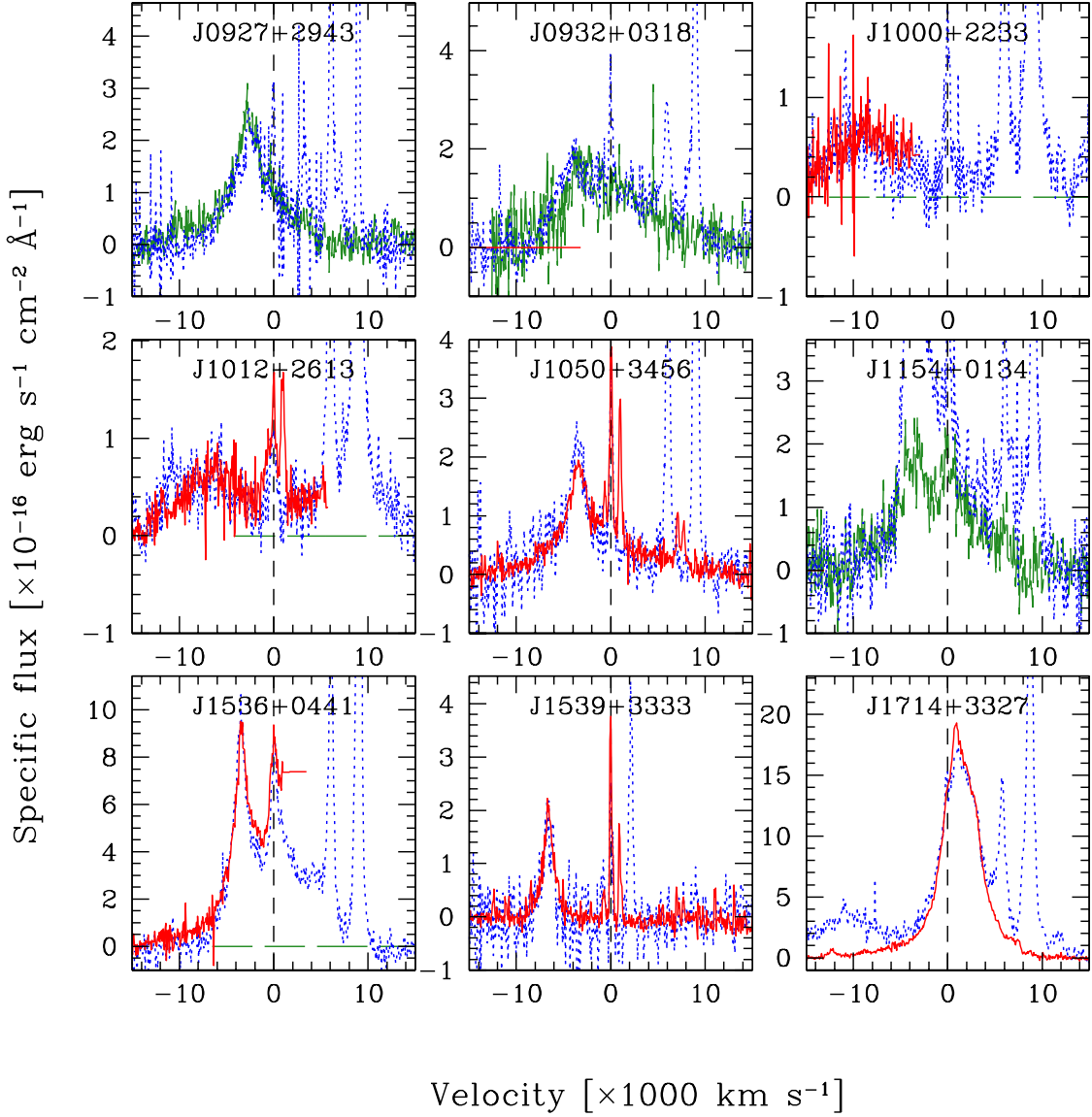}\\
\caption{Velocity diagrams of H$\alpha$ (red, solid lines), H$\beta$ 
  (blue, dotted lines) and Mg{\sc ii} (green, dashed lines) for all the 
  spectroscopically-identified MBHB candidates. The flux of H$\beta$ is 
  scaled up to match the one of H$\alpha$ or Mg{\sc ii}. Adapted from 
  \citet{vivi11}.}\label{bhb_vel}
\end{center}
\end{figure}

It should be noted that alternative interpretations for the spectral
properties of the known candidates are available:

\begin{itemize} 

\item[{\it i-}] Modest line shifts ($\lesssim$ 500 km s$^{-1}$) are
often observed in ``normal'' AGN \citep{bonning07}.

\item[{\it ii-}] Similarly, small velocity shifts ($<$4000 km
s$^{-1}$), can be associated to the remnant of a binary coalescence,
recoiling because of anisotropic gravitational wave emission
\citep{komossa08}\footnote{However, if the galaxy merger is gas rich,
the maximum recoil velocity is expected to be $< 100$ km s$^{−1}$
\citep{bogdanovic07, dotti10,
 volonteri10, kesden10}}.  

\item[{\it iii-}] An unobscured MBHB with both MBHs active could
resamble the spectrum of a double peaked emitter \citep[see,
e.g.,][]{eracleous94}, where broad doublepeaked lines are emitted
because of the almost edgeon, disklike structure of the broad line
region of a single MBH.

\item[{\it iv-}] Chance superposition of two AGN (or an AGN-galaxy
superposition) within the angular resolution of the used spectrograph
can also mimic velocity shifts of different line systems
\citep{heckman09}.

\end{itemize}

\noindent The simplest way to discriminate between these scenarios and the MBHB
hypothesis would be to look for a periodic oscillation of the broad
 line shifts around the host galaxy redshift. However, the orbital
 period of the binary could be too long to be easily observed
 \citep{begelman80}. Noticeably, \cite{eracleous11} observed a
 variation in the shifts at two different epochs in 14 out of 88
 candidates, with resulting accelerations between -120 and +120
 km/s/yr. Longer temporal baselines are needed to prove the MBH binary
 interpretation for these objects.

In order to increase the number of known MBHB candidates, and to
confirm their interpretation, it is therefore of fundamental
importance to identify new signatures of MBHBs. The simultaneous
observation of various MBHB signatures could represent the only way to
firmly validate the MBHB scenario in the known candidates. A
possibility is to look for peculiar flux ratios between high- and
low-ionization broad lines. The broad line region of each MBH in a
binary can be perturbed and disrupted by the gravitational potential
of the companion. External shells of the broad line region (where most
of the low-ionization line flux is emitted) are affected first,
resulting in peculiar flux ratios \citep{montuori10}. This criterion
is particularly interesting for quasars at high redshift ($z \sim 2$),
where high and low ionization lines are observable in large surveys as
the SDSS, while the most prominent narrow lines, needed to measure a
shift related to the orbital motions, are not present in the SDSS
spectra anymore.

At even closer separations between the two MBHs, when the size of the
BLR is significantly larger than the semimajor axis of the binary ($a
\lsim 0.01$ pc), the optical and UV spectral features discussed above
become more complex and not directly related with the period of the
binary \citep{shenloeb}. However, in this case typical MBHB periods
are $\lsim 10$ yr, opening the interesting opportunity of directly 
detect periodic variability of the system, related to the periodicity
of the accretion flows \citep{artymowicz96,hayasaki08,sesana11}. 
Such close separations are particularly interesting, since they will be 
proven by the ongoing and future pulsar timing arrays \citep[PTAs,][]{hobbs10}.
In this context, \cite{sesana11} estimate that up to few hundred MBHBs
contributing to the GW signal in the PTA band may be identified 
through their periodicity in future X-ray all sky surveys. Among those, 
few exceptionally bright sources may be resolved both in the GW and 
in the electromagnetic window through the detection of peculiar 
double K-$\alpha$ fluorescence lines, offering a unique multimessenger 
astronomy opportunity. An alternative possibility,
suggested by \cite{tanaka11}, is that the presence of a 
circumbinary cavity results in a suppression of the UV soft-X
emission. MBHBs close to coalescence may therefore be identified
as AGNs with exceptionally faint UV X-ray continuum.

\section{Conclusions}\label{conclusions}

We reviewed the most recent findings about the dynamical evolution of
MBHBs and their detectability. Regarding the binary dynamics, in the
last few years we recognized the importance of the medium/large scale
galactic structures ($\gsim 100$ pc) in the dynamical evolution of the
binaries. In gas free environments, the shape of the bulge potential
is directly related to the possibility of a MBHB to reach the final
coalescence.  In a spherical system, the stars that can interact with
the binary (i.e. that orbit within its loss--cone) are evacuated from
the center before the binary can coalesce \citep[e.g.][]{mm01}. A fast
refilling of the loss--cone, that can result in the merger of the two
MBHs, is possible in triaxial systems, in which the angular momentum
components of the orbiting stars are not conserved
\citep[e.g.][]{pm04}. Recently, thanks to the advent of GPU computing,
large scale galaxy mergers proved the occurrence of such a
replenishment in more realistic scenarios \citep{khan11, preto11,
  mg11}. The evolution of the binary eccentricity $e$ depends on the
dynamical properties of the core as well. If the MBHs are embedded in
a non rotating stellar system, the general trend is towards an
increase of $e$ with time \citep[e.g.][]{qui96}. This promotes the
coalescence of the MBHs, since gravitational wave emission is more
efficient for eccentric binaries. In rotating systems, on the other
hand, the evolution of $e$ depends on the orientation of the binary: a
binary co-rotating with the stellar cusp tends to decrease its
eccentricity, while in the counter--rotating case $e$ grows up to
$\gsim 0.95$ \citep{sgd11}. This clear cut scenario could be modified
by the interaction with stars on quasi--radial, centrophilic orbits. A
study of the orbital properties of these stars has not been presented
to date.

Galaxy mergers can easily promote strong inflows of gas towards the
center of the galaxy remnant\citep[e.g.][]{mayer07}. Hence, it is
fundamental to understand how the presence of massive gas structures
in the cores modify the dynamical evolution of the forming MBHBs.  It
has been proven that the interaction with circumnuclear ($\sim 100$
pc) disks can result in a fast ($\lsim 10^7$ yr) formation of a MBHB
\citep[e.g.][]{escala05}. After this fast transient, the binary is
thought to open a gap in the central gas distribution
\citep[e.g.][]{linpapaloizou79a}, and any further evolution is
mediated by the interaction between the MBHs and the inner edge of the
circumbinary disk.

Simulations and analytical studies about the interaction of MBHBs and
circumbinary disks have improved our knowledge of the physical
processes in play, and their effect onto the binary
evolution. Howerver, a complete understanding of the binary/disk
interaction is still to come. For example, in many investigations the
circumbinary disk is assumed to be cylindrically symmetric, i.e., the
study is reduced to an effective one-dimensional problem. The
assumption of cylindrical symmetry removes any possibility of studying
the effects of structures in the disk, and, most importantly, of gas
streams periodically inflowing from the disk onto the two MBHs,
routinely observed in simulations
\citep[e.g.][]{artymowicz96,hayasaki08,cuadra09,roedig11,sesana11}. The
torques exerted by these inflowing streams has not been studied in
detail yet \citep[with the notable exception of][]{macf08}, and could
provide additional help (or resistance) in bringing the binary to the
final coalescence.

The final fate of a binary embedded in such a circumbinary disk is
still debated. If the disk is continously refueled from any
larger-scale gas distribution, a fast coalescence can easely be
achieved. However, if the binary cannot interact with enough gas
(e.g. because it turns in stars), the circumbinary disk gets evacuated
and fails in bringing the MBHs to coalescence
\citep[e.g.][]{lodato09}. As a consequence, as in the stellar
scenario, the final fate of the binary depends on the properties of
larger scale structures, and its abitily to efficiently refuel the
proximity of the MBHB with fresh gas. In principle, in presence of an
intense inflow toward the center, the binary could fail in opening a
gap at all, and the interaction between MBHs and a closer/denser gas
structure could result in a faster coalescence of the MBHB
\citep{escala05}. Furthermore, if the angular momentum of gas can be
efficiently re-shuffled, inflowing streams could form counter-rotating
circumbinary disks, that can also promote the binary coalescence
\citep{nixon11}.  Only recently the formation of a gap has been
observed in large scale simulations \citep{escala05,dotti09}, in which
the evolution of an extended ($\sim$ 100 pc) massive disk is followed
(as massive as the binary in Escala et al., up to $\gsim 10 M_{\rm BHB}$ in
Dotti et al., see figure~\ref{gap}). The spatial and mass resolution
of these simulations do not allow yet a detailed study of the sub-pc
evolution of the binary, down to a possible coalescence. Simulating
the evolution of a sub-pc binary starting from large scale initial
conditions, that can constrain the properties of the nuclear inflows
together with the evolution of the binary, is the fundamental
improvement needed to build a coherent picture of MBHBs in gas rich
environment.

To summarize the recent findings present in literature, we can draw a
comparison between orbital decay timescales obtained considering
different scenarios, for equal mass binaries.
\begin{itemize} 
\item {\it In dense stellar environments:} {\it if} the loss-cone of the
  MBHB is constantly refilled (see Section~2.1), a binary
  of $10^6 \msun$ with an initial separation of $a_0 \approx 1$ pc will 
  coalesce in $\approx 3\times10^7-10^8$ years , while a
  binary 100 times more massive will inspiral for about  
  $\approx 10^8-10^9$ years before reaching the
  final coalescence \citep{sesana10,preto11}.
\item {\it in gas rich environments:} {\it if} the interaction with a
  steady, long-lived, {\it corotating}, maximally massive circumbinary
  disk is responsable for the MBHB orbital decay (see Section~3.2),
  the orbital decay timescale
  for a $10^6 \msun$ equal mass binary is less than the age of the
  Universe if the binary starts at an initial separation $a_0 \approx
  0.03$ pc ($\sim 10$ times smaller than the separation at which the
  MBHs bind in a binary $a_{\rm BHB}$), and is 2 orders of magnitude
  shorter for $a_0 \approx 0.01$ pc \citep{cuadra09}. 
    For a $10^8 \msun$ equal mass binary, this timescale is less than
    the age of the Universe if $a_0 \approx 0.05$ pc \citep{cuadra09},
    $\sim 100$ times less than $a_{\rm BHB}$.
\item {\it in gas rich environments:} {\it if} the MBHB interacts with
  a continuous sequence of counterrotating accretion disks with an
  accretion rate corresponding to the MBH Eddington limit, the orbital
  decay timescale is of the order of $10^8$ yr, regardless the binary
  mass, for $a_0 < a_{\rm BHB}$. In this case, the coalescence
  timescale increases linearly with the inverse of the accretion rate.
\end{itemize}

Note that the estimates in gas rich environment should be considered
as lower limits, since they assume continuous accretion and, as
stressed above, a continuous re-fuelling of the disks from larger
scales. A single, not re-fuelled disk with enough mass and angular
momentum to bring the MBHs to the final coalescence form $a_0\lsim
0.1$ pc would undergo fragmentation and star formation, as discussed
in \cite{lodato09}.

The presence of gas close to the binary is necessary to its
detection. If at least one of the two MBHs is active, the orbital
velocity of the binary can result in a frequency shift between the
broad emission lines and the narrow emission lines
\citep[e.g.][]{begelman80}.  This shift is expected to change
periodically, on the orbital period.  Using this technique few tens of
MBHB candidates have been selected \citep[e.g.][]{vivi11,
  eracleous11}.  For all the candidates discussed to date, possible
explanations other than a MBHB have been proposed (see
section~\ref{roris}). Moreover, the orbital period expected for such
binaries is often too long to be observed, thus the periodic variation
of the velocity shift cannot be used (yet) to determine the real
nature of the candidates. It is therefore fundamental to couple this
with other (independent) signatures of MBHBs, to confirm their nature. 

In the near future, space based interferometers like NGO will detect
the GWs emitted in the late inspiral and final coalescence of MBHBs,
opening the fascinating prospective of multimessenger astronomy. The
identification of an electromagnetic counterpart to a gravitational
wave detection of merging MBHs can teach us about relativity,
accretion physics, galaxy formation and evolution and, more in
general, cosmology. A large variety of potential EM signatures have
been proposed \citep[see, e.g.][and references therein]{schnittman11}.
In particular, thanks to the recent quick progresses in numerical
relativity, simulations of the coalescence process in presence of
matter is now becoming possible
\citep{bode10,palenzuela10,farris11,moesta11}, and in the coming years
may provide valuable insights about the nature of the associated
electromagnetic signals.  However, most of the counterparts discussed
to date depend on some assumptions on the distribution of gas and/or
stars in the immediate vicinities of the MBHB. A theoretical
comprehension of the dynamics of matter close to the binary, necessary
to understand the fate of a MBHB, is also fundamental to constrain the
observability of an electromagnetic counterpart.

Tracing the formation, evolution and final fate of MBHBs is certainly
one of the open challenges of contemporary astrophysics.  A better
understanding of the interaction between MBHs and gas, and the
prediction of new peculiar observational features of MBHBs are needed
to unambiguosly constrain their properties and demography, adding an
important missing piece to the galaxy evolution puzzle.

\acknowledgments

We are grateful to Jorge Cuadra, Monica Colpi and
Paraskevi Tsalmantza for useful discussions and comments.

\clearpage

\end{document}